\documentclass[a4paper,11pt]{article}
\pdfoutput=1
\usepackage{jheppub}
\usepackage[utf8]{inputenc}
\usepackage{amsmath}
\usepackage{amsfonts}
\usepackage{amssymb}
\usepackage{physics}
\usepackage{tikz}
\usepackage{appendix}
\usepackage{hyperref}
\usepackage{xcolor}
\usepackage{graphicx}
\usepackage{float}

\newcommand{\fft}[2]{\frac{#1}{#2}}

\newcommand{\nn}{\nonumber}

\newcommand{\cN}{\mathcal{N}}
\newcommand{\cL}{\mathcal{L}}

\newcommand{\cO}{\mathcal{O}}
\newcommand{\cC}{\mathcal{C}}

\newcommand{\cF}{\mathcal{F}}
\newcommand{\cS}{\mathcal{S}}
\newcommand{\bR}{\mathbb{R}}
\newcommand{\cR}{\mathcal{R}}

\numberwithin{equation}{section}

\begin{document}
\preprint{LCTP-21-38}

\title{Four-derivative Corrections to Minimal Gauged Supergravity in Five Dimensions}
\author{James T. Liu and Robert J. Saskowski}

\affiliation{Leinweber Center for Theoretical Physics, Randall Laboratory of Physics\\The University of Michigan, Ann Arbor, MI 48109-1040, USA }

\emailAdd{jimliu@umich.edu, rsaskows@umich.edu}

\abstract{
    We study four-derivative corrections to pure $\cN=2$, $D=5$ gauged supergravity. In particular, we find that, up to field redefinitions, there is a single four-derivative superinvariant that one can add to the action, up to factors of the two-derivative action. Consequently, this selects out a unique set of coefficients for the four-derivative corrections. We confirm these coefficients (in the ungauged limit) on the BMPV solution.
}

\maketitle

\section{Introduction}
In the last two decades, the AdS/CFT correspondence has led to a number of remarkable results. In particular, we have a dictionary between gravitational theories in $(d+1)$-dimensional AdS spacetimes and superconformal field theories living on the $d$-dimensional boundary. The gravity side is often described by an effective supergravity theory, supplemented by higher-derivative corrections coming from stringy $\alpha'$ corrections to the effective action.

While such corrections can be obtained directly from the underlying string theory, they can also be parametrized more generally by working directly in the supergravity theory.  Various higher derivative superinvariants have been constructed, both in the off-shell conformal supergravity approach and in the Poincar\'e frame.  The former is particularly powerful, although it is generally limited to theories with at most eight real supercharges.  In this manner, four-derivative corrections have been constructed in four-dimensional $\mathcal N=2$ \cite{deWit:1979dzm,deWit:1980lyi,deWit:1984rvr,deWit:1984wbb,LopesCardoso:2000qm,deWit:2006gn,Butter:2013lta}, five-dimensional $\mathcal N=2$ \cite{Hanaki:2006pj,Bergshoeff:2011xn,Ozkan:2013uk,Ozkan:2013nwa,Baggio:2014iv}, and six-dimensional $\mathcal N=(1,0)$ \cite{Bergshoeff:1986vy,Bergshoeff:1986wc,Bergshoeff:1987rb,Novak:2017wqc,Butter:2018wss} supergravities.  As these constructions are not yet in the Poincar\'e frame, an additional step is needed in integrating out the auxiliary fields in order to obtain conformally gauge-fixed superinvariants.

Supersymmetric higher derivative actions have a wide range of applications from black holes to precision holography.  Higher derivative corrected black holes provide a window into quantum gravity and can also shed light on the black hole weak gravity conjecture and the swampland.  Higher derivative corrections have also played an important role in holographic hydrodynamics and bounds on $\eta/s$, the ratio of the shear viscosity to the entropy density of the plasma.

Recently, there has been interesting work on four-derivative corrections in $\cN=2$, $D=4$ gauged supergravity with regards to AdS$_4$/CFT$_3$ holography \cite{Bobev:2020egg,Bobev:2020zov, Bobev:2021oku}. In particular, there are two off-shell four-derivative superinvariants that one can add to the action, namely the Weyl multiplet and the $\mathbb{T}$log multiplet
\begin{equation}
    S_\text{HD}^{4d}=S_{2\partial}+\alpha_1 S_{\text{Weyl}}+\alpha_2 S_{\mathbb{T}\text{log}},
\end{equation}
where $S_{2\partial}$ denotes the usual two-derivative action. The authors of \cite{Bobev:2020egg,Bobev:2021oku} showed that on-shell these reduce to a single four-derivative superinvariant, the Gauss-Bonnet action, as well as a term proportional to the two derivative action
\begin{equation}
    I_\text{HD}^{4d}= \qty(1+\frac{4}{L^2}(\alpha_2-\alpha_1))I_{2\partial}+\alpha_1I_{\text{GB}},
\label{eq:SHDos}
\end{equation}
where $L$ is the four-dimensional AdS radius.  Here $I$ denotes the on-shell value of the action~$S$.

At this point, a comment on our usage of off-shell and on-shell is in order, as there are perhaps two notions of on/off-shell in the supergravity literature: on-shell in the context of conformal supergravity and on-shell in the context of evaluating actions.  In the former, the off-shell action includes non-dynamical auxiliary fields needed for off-shell closure of the supersymmetry algebra, while on-shell indicates that the auxiliary fields have been integrated out.  In the later, on-shell means the equations of motion have been used when evaluating the action, thus yielding an on-shell value that can no longer be used for dynamics.  In the interest of clarity, we will refer to on-shell in the conformal supergravity sense as the Poincar\'e frame, and reserve the use of on-shell to denote computing the on-shell value of the action%
\footnote{On-shell here should not be confused with the use of field redefinitions to transform the higher-derivative actions.  As further discussed in Section~\ref{sec:fieldredef}, we can shift the higher derivative action by terms proportional to the two-derivative equations of motion.  However, the resulting action is still a dynamical action that is physically equivalent to the original one in that it yields identical on-shell observables.}.

The remarkable feature of the four-derivative on-shell action, (\ref{eq:SHDos}), as applied to AdS$_4$/CFT$_3$, is that is provides a natural split between geometrical and theory dependent parameters.  The former arise from the on-shell values of $I_{2\partial}$ and $I_{\mathrm{GB}}$, while the latter correspond to $\alpha_1$, $\alpha_2$ and the five-dimensional AdS radius, $L$.  As shown in \cite{Bobev:2020egg}, the partition function of the dual SCFT obtained from (\ref{eq:SHDos}) thus takes the universal form
\begin{equation}
    -\log Z=\pi\cF\qty[AN^{3/2}+BN^{1/2}]-\pi(\cF-\chi)CN^{1/2}+o(N^{1/2}),
\end{equation}
where $\cF$ and $\chi$ depend on the three-dimensional manifold that the SCFT lives on while $\{A,B,C\}$ are related to $\{\alpha_1,\alpha_2,L\}$, and are specific to the SCFT in question. This motivates us to consider the analogous case for AdS$_5$/CFT$_4$, namely $\cN=2$, $D=5$ gauged supergravity.

In the five-dimensional case, three independent four-derivative off-shell terms have been constructed, corresponding to the supersymmetrization of $R_{\mu\nu\rho\sigma}^2$, $R_{\mu\nu}^2$ and $R^2$ \cite{Ozkan:2013nwa}%
\footnote{The supersymmetrization of the 5D Weyl-squared action was performed in \cite{Hanaki:2006pj}, and the resulting corrections to black holes have been investigated in \cite{Castro:2007ci,Castro:2007hc,Castro:2007sd,Castro:2008ne,Castro:2008ys,Alishahiha:2007nn}.}.
Following \cite{Ozkan:2013nwa}, we choose an equivalent basis of $C_{\mu\nu\rho\sigma}C^{\mu\nu\rho\sigma}+\frac{1}{6}R^2$, $R_{\mu\nu\rho\sigma}R^{\mu\nu\rho\sigma}$, and $R^2$.  Then
\begin{equation}
    S_\text{HD}^{5d}=S_{2\partial}+\alpha_1 S_{C^2+\frac{1}{6}R^2}+\alpha_2 S_{(\text{Riem})^2}+\alpha_3 S_{R^2}.
\end{equation}
After some work, we find a direct analogue of the four-dimensional result for the on-shell value of the action:
\begin{equation}
    I_\text{HD}^{5d}=\qty(1+\frac{8\alpha_1-24\alpha_2-205\alpha_3}{2L^2})I_{2\partial}+(\alpha_1-2\alpha_2)I_\text{GB}^\text{susy},
\label{eq:onshellAction}
\end{equation}
where $S_{\text{GB}}^\text{susy}$ is the supersymmetrization of the Gauss-Bonnet action in 5D and $I_{\text{GB}}^\text{susy}$ is its on-shell value. However, this is slightly more complicated than the 4D case, as the Gauss-Bonnet action is no longer topological in 5D, and, as noted in \cite{Bobev:2021qxx}, the two-derivative solutions no longer satisfy the four-derivative equations of motion as was shown in the 4D case \cite{Bobev:2020egg,Bobev:2021oku}.

One important aspect of \eqref{eq:onshellAction} is that, up to field redefinitions and an overall coefficient, the four-derivative action $S_{\mathrm{GB}}^\text{susy}$ is completely fixed.  As shown in \cite{Myers:2009ij}, five-dimensional Einstein-Maxwell theory admits five independent four-derivative terms up to field redefinitions.  Here we choose a somewhat different but equally valid parametrization from that of \cite{Myers:2009ij}
\begin{equation}
    e^{-1}\cL_{4\partial}=c_1\mathcal I_\text{GB}+c_2 C_{\mu\nu\rho\sigma}F^{\mu\nu}F^{\rho\sigma}+c_3 (F^2)^2+c_4 F^4+c_5\epsilon^{\mu\nu\rho\sigma\lambda}R_{\mu\nu ab}R_{\rho\sigma}{}^{ab}A_{\lambda}.
\label{eq:param4d}
\end{equation}
The fields are normalized according to the two-derivative Lagrangian
\begin{equation}
    e^{-1}\cL_{2\partial}=R+12g^2-\fft14F^2-\fft1{12\sqrt3}\epsilon^{\mu\nu\rho\sigma\lambda}F_{\mu\nu}F_{\rho\sigma}A_\lambda.
\end{equation}
Here $g=1/L$ is the gauge coupling constant, $A_\mu$ is the graviphoton field with $F=\dd{A}$ its corresponding field strength, and $\mathcal I_\text{GB}=R_{\mu\nu\rho\sigma}^2-4R_{\mu\nu}^2+R^2$ is the usual Gauss-Bonnet combination. Our notational convention, here and throughout this paper, is that $F^2=F_{\mu\nu}F^{\mu\nu}$ and $F^4=F_{\mu\nu}F^{\nu\rho}F_{\rho\sigma}F^{\sigma\mu}$. For the supersymmetric four-derivative invariant, $S_{\mathrm{GB}}$, the above result, \eqref{eq:onshellAction}, fixes these coefficients up to an overall factor
\begin{equation}
    c_1=-2 c_2=8 c_4=2\sqrt{3} c_5,\qquad
    c_3=0.
\label{eq:correctc}
\end{equation}

Although we refer to $S_{\mathrm{GB}}^\text{susy}$ as the supersymmetric Gauss-Bonnet action in 5D, it could equivalently have arisen from integrating out the auxiliary fields of the off-shell Weyl-squared action \cite{Hanaki:2006pj}.  This was done earlier in \cite{Cremonini:2008tw,Myers:2009ij}, which however led to a different set of coefficients \cite{Cremonini:2020smy}
\begin{equation}
    \tilde c_1 = -2\tilde c_2 = -6\tilde c_3 = - \frac{24}{11}\tilde c_4 = 2\sqrt{3}\tilde c_5.
\label{eq:wrongc}
\end{equation}
We resolve this conflict by checking the four-derivative correction to the supersymmetric BMPV solution \cite{Breckenridge:1996mpv}, which corresponds to the ungauged limit ($L\to\infty$) of (\ref{eq:onshellAction}).  In particular, we show that the BPS condition, $M=\fft{\sqrt3}2|Q|$, only remains satisfied for the present choice of coefficients, (\ref{eq:correctc}).  We will comment further on this discrepancy below.

This note is organized as follows. In Section~\ref{sec:sugra}, we show that the three off-shell four-derivative superinvariants reduce to a single Poincar\'e frame invariant, up to field redefinitions and factors of the two-derivative action. We then proceed in Section~\ref{sec:BMPV} to check our results on the BMPV black hole solution. Finally, we conclude in Section \ref{sec:discussion} with additional comments on resolving the discrepancy between (\ref{eq:correctc}) and (\ref{eq:wrongc}) and with some open questions. 

While this work was being completed, we became aware of \cite{Bobev:2021qxx}, which overlaps with some of our results. However, it should be noted that if we transform the results of \cite{Bobev:2021qxx} into the field-redefinition frame (\ref{eq:param4d}), we get exactly the result (\ref{eq:correctc}).

\section{Higher-derivative supergravity}\label{sec:sugra}
Minimal $D = 5$ gauged supergravity has a single symplectic Majorana supercharge. The field content is then just the $\cN=2$ gravity multiplet $(e_\mu^a,\psi_\mu,A_\mu)$. The two-derivative (bosonic) Lagrangian in the Poincar\'e frame is given by
\begin{equation}
    S_{2\partial}=\int\qty[(R+12 g^2)\star1 -\frac{1}{2}F\land \star F -\frac{1}{3\sqrt{3}}F\land F\land A],
\label{eq:2derivAction}
\end{equation}
where $R$ is the Ricci scalar, $F = \dd{A}$ is the field strength of the U(1) graviphoton, and $g=1/L$ is the $U(1)$ gauge coupling that may be identified as the inverse AdS radius. Note that we choose to work in conventions such that $16\pi G_N\equiv 1$, the metric has signature $(-,+,+,+,+)$, and $R_{\mu\nu}=R^{\rho}{}_{\mu\rho\nu}$. The two-derivative equations of motion are
\begin{subequations}
\begin{align}
    0=\mathcal E^\mu&\equiv\nabla_\nu F^{\nu\mu}+\frac{1}{2\sqrt{6}}\epsilon^{\mu\nu\rho\sigma\lambda}F_{\nu\rho}F_{\sigma\lambda},\\
    0=\mathcal E_{\mu\nu}&\equiv R_{\mu\nu}-\left(F_\mu^{\ \,\sigma}F_{\nu\sigma}-\frac{1}{6}g_{\mu\nu}F^2-4g^2 g_{\mu\nu}\right).
\end{align}
\label{eq:eoms}
\end{subequations}

At the four-derivative level, there are three terms that can be added to the action, corresponding to the supersymmetrizations of $(R_{\mu\nu\rho\sigma})^2$, $(R_{\mu\nu})^2$, and $R^2$. However, we choose to parametrize these as $(C_{\mu\nu\rho\sigma})^2+\tfrac{1}{6}R^2$, $(R_{\mu\nu\rho\sigma})^2$, and $R^2$, as the supersymmetrizations of these combinations have been found in \cite{Ozkan:2013nwa, Ozkan:2013uk} via conformal supergravity methods.  We consider these three off-shell invariants below and perform the field redefinitions necessary to transform them into the parametrization of (\ref{eq:param4d}).

\subsection{The action corresponding to $(C_{\mu\nu\rho\sigma})^2+\tfrac{1}{6}R^2$}
\label{sec:fieldredef}

The supersymmetrization of the square of the Weyl tensor was originally considered in \cite{Hanaki:2006pj} using the standard Weyl multiplet, and subsequently in \cite{Ozkan:2013nwa,Ozkan:2013uk} using the dilaton Weyl multiplet.  In the latter case, the supersymmetric completion of $C_{\mu\nu\rho\sigma}^2$ picks up an additional $\fft16R^2$ term, and the Poincar\'e frame action takes the form \cite{Ozkan:2013nwa,Ozkan:2013uk}%
\footnote{Note that our conventions are $16\pi G_N=1$, whereas the conventions in \cite{Ozkan:2013nwa,Ozkan:2013uk} are that $8\pi G_N=1$.}
\begin{align}
    e^{-1}\cL_{C^2+\frac{1}{6}R^2}&=\frac{1}{4} R_{\mu \nu \rho \sigma} R^{\mu \nu \rho \sigma}- R_{\mu \nu} R^{\mu \nu}+\frac{1}{24} R^{2}+\frac{128}{3} D^{2}+\frac{1}{8} \epsilon_{\mu \nu \rho \sigma \lambda} C^{\mu} R^{\nu \rho \tau \delta} R^{\sigma \lambda}_{\ \ \ \tau \delta}\nn \\
    &\quad-\frac{32}{3} R_{\mu \nu \rho \sigma} T^{\mu \nu} T^{\rho \sigma}+4 R_{\mu \nu \rho \sigma} G^{\mu \nu} T^{\rho \sigma}+\frac{2}{3}R T_{\mu \nu} G^{\mu \nu}-\frac{16}{3} R_{\mu \nu} G_{\sigma}^{\ \,\mu} T^{\sigma \nu}\nn \\
    &\quad-\frac{128}{3} R^{\mu \nu} T_{\sigma \mu} T_{\ \,\nu}^{\sigma}+\frac{16}{3} R T^{2}-\frac{64}{3} D T_{\mu \nu} G^{\mu \nu}+\frac{2048}{9} T^{2} D-\frac{128}{3} \nabla_{\mu} T_{\nu \rho} \nabla^{\mu} T^{\nu \rho}\nn \\
    &\quad+\frac{128}{3} \nabla^{\mu} T^{\nu \rho} \nabla_{\nu} T_{\mu \rho}-\frac{256}{3} T_{\mu \nu} \nabla^{\nu} \nabla_{\sigma} T^{\mu \sigma}+2048 T^{4}-\frac{5632}{27}\left(T^{2}\right)^{2}\nn\\
    &\quad-\frac{128}{9} T_{\mu \nu} G^{\mu \nu} T^{2}-\frac{512}{3} T_{\mu \sigma} T^{\sigma \rho} T_{\rho \nu} G^{\nu \mu}-\frac{256}{3} \epsilon_{\mu \nu \rho \sigma \lambda} T^{\mu \nu} T^{\rho \sigma} \nabla_{\tau} T^{\lambda \tau}\nn\\
    &\quad-\frac{64}{3} \epsilon_{\mu \nu \rho \sigma \lambda} G^{\mu \nu} T^{\rho \tau} \nabla_{\tau} T^{\sigma \lambda}-32 \epsilon_{\mu \nu \rho \sigma \lambda} G^{\mu \nu} T^{\rho}{ }_{\tau} \nabla^{\sigma} T^{\lambda \tau},\label{eq:Weyl2initial}
\end{align}
where $G=\dd{C}$, and we have relations
\begin{subequations}
\begin{align}
    C_{\mu}&=\sqrt{\frac{2}{3}}A_{\mu},\\
    T_{\mu\nu}&=\frac{3}{16}G_{\mu\nu}=\frac{1}{8}\sqrt{\frac{3}{2}}F_{\mu\nu},\\
    D&=-\frac{1}{32}R-\frac{1}{16}G^2-\frac{26}{3}T^2+2T^{\mu\nu}G_{\mu\nu}=-\frac{1}{32}R+\frac{1}{192}F^2.
\end{align}
\label{eq:identities1}%
\end{subequations}
We note that \cite{Ozkan:2013nwa,Ozkan:2013uk} derived the above action in the context of asymptotically Minkowski space; however, moving to the AdS case (or, equivalently, gauging the supergravity) does not affect \eqref{eq:Weyl2initial}. Rather, this gauging will only affect the two-derivative action by turning on a non-zero gauge parameter $g$, which does affect the field redefinitions used in the simplifications that follow. Making use of \eqref{eq:identities1}, one can simplify \eqref{eq:Weyl2initial} to
\begin{align}
    e^{-1}\cL_{C^2+\frac{1}{6}R^2}&=\frac{1}{4} R_{\mu \nu \rho \sigma} R^{\mu \nu \rho \sigma}- R_{\mu \nu} R^{\mu \nu}+\frac{1}{12} R^{2} +\frac{1}{4}R_{\mu\nu\rho\sigma}F^{\mu\nu}F^{\rho\sigma}-\frac{61}{432}(F^2)^2+\frac{5}{8}F^4\nn \\
    &\quad+\frac{1}{4\sqrt{6}}\epsilon_{\mu\nu\rho\sigma\lambda}A^\mu R^{\nu\rho\tau\delta}R^{\sigma\lambda}{}_{\tau\delta}+\frac{1}{9}RF^2-\frac{5}{3}R^{\mu\nu}F_{\sigma\mu}F^\sigma{}_\nu-(\nabla F)^2\nn\\
    &\quad+3(\nabla_\sigma F^{\mu\sigma})^2-F^{\nu\rho}[\nabla_\mu,\nabla_\nu]F^\mu{}_\rho+\frac{1}{4\sqrt{6}}\epsilon_{\mu\nu\rho\sigma\lambda}F^{\mu\nu}F^{\rho\sigma}\nabla_\tau F^{\tau\lambda}\nn \\
    &\quad-\frac{1}{2}\sqrt{\frac{3}{2}}\epsilon_{\mu\nu\rho\sigma\lambda}F^{\mu\nu}F^{\rho}{}_{\tau}\nabla^\sigma F^{\lambda\tau}.
\label{eq:C2R2}
\end{align}

We now perform a set of field redefinitions to put the action, (\ref{eq:C2R2}), into the canonical form (\ref{eq:param4d}).  Our starting point is a Lagrangian of the form
\begin{equation}
    \cL=\cL_{2\partial}+\alpha\cL_{4\partial},
\label{eq:L24}
\end{equation}
where $\cL_{2\partial}$ is given in (\ref{eq:2derivAction}) and we have introduced a parameter $\alpha$ to keep track of the derivative expansion.  Now consider a field redefinition
\begin{equation}
    g_{\mu\nu}\to g_{\mu\nu}+\alpha\delta g_{\mu\nu},\qquad A_\mu\to A_\mu+\alpha\delta A_\mu.
\label{eq:dgdA}
\end{equation}
Applying this to the full Lagrangian, (\ref{eq:L24}), and allowing for integration by parts results in
\begin{equation}
    e^{-1}\mathcal L\to e^{-1}\mathcal L_{2\partial}+\alpha\left(e^{-1}\mathcal L_{4\partial}+\mathcal E_{\mu\nu}\delta g^{\mu\nu}+\mathcal E^\mu\delta A_\mu\right)+\mathcal O(\alpha^2),
\label{eq:fredef}
\end{equation}
where the two-derivative Einstein and Maxwell equations, $\mathcal E_{\mu\nu}$ and $\mathcal E_\mu$, are given in (\ref{eq:eoms}).  Since we are only interested in the four-derivative effective action, we ignore all terms of $\mathcal O(\alpha^2)$ and higher.  By judicial choice of $\delta g_{\mu\nu}$ and $\delta A_\mu$, we are then able to transform the four-derivative Lagrangian into the form (\ref{eq:param4d}).

As seen in (\ref{eq:fredef}), field redefinitions allow us to shift the four-derivative action by terms proportional to the two-derivative equations of motion.  In practice, this means we can substitute the two-derivative equations of motion, (\ref{eq:eoms}), into the four-derivative action, (\ref{eq:C2R2}) to transform it into canonical form.  It is important to note, however, that we are treating the field redefinition, (\ref{eq:dgdA}), perturbatively in the derivative expansion.  In particular, while we are only considering the four-derivative terms, an infinite set of terms beyond four derivatives will be generated by the field redefinition.  Furthermore, a field redefinition in the path integral will also transform the measure and couplings to external sources.  Nevertheless, physical (on-shell) quantities computed before and after the field redefinition will remain unchanged at the four-derivative order.

With the above in mind, we now use integration by parts and the two-derivative equations of motion, \eqref{eq:eoms}, to make the replacements
\begin{subequations}
\begin{align}
    (\nabla F)^2&\to-\frac{1}{3}(F^2)^2-\frac{2}{3}F^4+R_{\mu\nu\rho\sigma}F^{\mu\nu}F^{\rho\sigma}+8g^2 F^2,\\
    (\nabla_\mu F^{\mu\nu})^2 &\to -\frac{1}{3}(F^2)^2+\frac{2}{3}F^4,\\
    \epsilon_{\mu\nu\rho\sigma\lambda}F^{\mu\nu}F^{\rho\sigma}\nabla_\tau F^{\tau\lambda}&\to\frac{4}{\sqrt{6}}\qty[(F^2)^2-2F^4],\\
    F^{\nu\rho}[\nabla_\mu,\nabla_\nu]F^\mu{}_\rho&\to F^4-\frac{1}{6}(F^2)^2-\frac{1}{2}R_{\mu\nu\rho\sigma}F^{\mu\nu}F^{\rho\sigma}-4g^2F^2,
\end{align}
\label{eq:simplifications1}%
\end{subequations}
inside the four-derivative action. These replacement rules are proved in Appendix \ref{app:simplifications}. Using \eqref{eq:eoms} and \eqref{eq:simplifications1}, we find that \eqref{eq:Weyl2initial} can be reduced to the simple form
\begin{align}
    e^{-1}\cL_{C^2+\frac{1}{6}R^2}&=\frac{1}{4}\mathcal I_{\text{GB}}-\frac{1}{4}R_{\mu\nu\rho\sigma}F^{\mu\nu}F^{\rho\sigma}-\frac{1}{16}(F^2)^2+\frac{11}{24}F^4+\frac{1}{4\sqrt{6}}\epsilon^{\mu\nu\rho\sigma\lambda}R_{\mu\nu ab}R_{\rho\sigma}{}^{ab}A_\lambda\nn\\
    &\quad+\frac{2}{3}g^2 F^2+20g^4,
\label{eq:invariant1}
\end{align}
where $\mathcal I_\text{GB}$ is the usual Gauss-Bonnet combination
\begin{equation}
    \mathcal I_\text{GB}=(R_{\mu\nu\rho\sigma})^2-4(R_{\mu\nu})^2+R^2.
\end{equation}

\subsection{The action corresponding to $(R_{\mu\nu\rho\sigma})^2$}

We now turn to the supersymmetrization of $(R_{\mu\nu\rho\sigma})^2$, which was found in \cite{Ozkan:2013nwa,Ozkan:2013uk} to be
\begin{align}
    e^{-1}\cL_{(\text{Riem})^2}&=-\frac{1}{2}\left(R_{\mu \nu a b}\left(\omega_{+}\right)-G_{\mu \nu} G_{a b}\right)\left(R^{\mu \nu a b}\left(\omega_{+}\right)-G^{\mu \nu} G^{a b}\right)\nn \\
    &\quad-\frac{1}{4} \epsilon^{\mu \nu \rho \sigma \lambda}\left(R_{\mu \nu a b}\left(\omega_{+}\right)-G_{\mu \nu} G_{a b}\right)\left(R_{\rho \sigma}^{\ \ \ a b}\left(\omega_{+}\right)-G_{\rho \sigma} G^{a b}\right) C_{\lambda}\nn \\
    &\quad-\epsilon^{\mu \nu \rho \sigma \lambda} B_{\rho \sigma}\left(R_{\mu \nu a b}\left(\omega_{+}\right)-G_{\mu \nu} G_{a b}\right) \nabla_{\lambda}\left(\omega_{+}\right) G^{a b}\nn\\
    &\quad-\nabla_{\mu}\left(\omega_{+}\right) G^{a b} \nabla^{\mu}\left(\omega_{+}\right) G_{a b},
\end{align}
where $H=\dd{B}+\tfrac{1}{2}C\land G$ and we have
\begin{subequations}
\begin{align}
    H_{\mu}^{\ \,ab}&=-\frac{1}{4}e_{f\mu}\epsilon^{fabcd}G_{cd},\\
    \omega_{+\mu}^{\ \ ab}&=\omega_\mu^{ab}+H_\mu^{\ \,ab},
\end{align}
\label{eq:torsionful}%
\end{subequations}
where $\omega^{ab}$ is the torsion-free spin connection. Making use of the standard formula
\begin{equation}
    R^{ab}(\omega_+)=\dd{\omega_+^{ab}}+\omega_+^{ac}\land \omega_{+c}{}^b,
\end{equation}
we may rewrite $R_{\mu\nu ab}$ in a manifestly torsion-free way
\begin{align}
    R^{ab}_{\mu\nu}(\omega_+)&=R^{ab}_{\mu\nu}+\frac{1}{2}\epsilon^{fabcd}e_{f[\mu}\nabla_{\nu]}G_{cd}\nn\\
    &\quad+\frac{1}{4}\qty(2G_{[\mu}{}^{a}G_{\nu]}^b+2G^{a\delta}G_{\delta[\mu}e^b_{\nu]}-2G^{b\delta}G_{\delta[\mu}e^a_{\nu]}+G^2e^a_{[\mu}e^b_{\nu]}).
\end{align}
Using \eqref{eq:torsionful}, we also see that
\begin{equation}
    \nabla_\mu(\omega_+)G_{ab}=\nabla_\mu G_{ab}-\frac{1}{2}e^{f}_\mu\epsilon_{f[a|ced}G^{ed}G^c{}_{|b]}.
\end{equation}

It is now straightforward to work out that
\begin{align}
    \left(R_{\mu \nu a b}\bigl(\omega_{+}\right)-G_{\mu \nu} G_{a b}\bigr)^2=&(R_{\mu \nu a b})^2-\frac{3}{2}R^{\mu\nu\rho\sigma}G_{\mu\nu}G_{\rho\sigma}-2R_{\mu\nu}G^{\mu\sigma}G^\nu{}_\sigma\nn\\
    &+\frac{1}{2}G^2R-(\nabla G)^2-(\nabla_\mu G^{\mu\nu})^2+\frac{5}{8}\qty(G^2)^2+\frac{9}{8}G^4\nn\\
    &-\frac{3}{4}\epsilon^{\mu\nu\rho\sigma\lambda}G_{\mu\nu}G_{\rho\sigma}\nabla^\tau G_{\tau\lambda}.
\end{align}
By using some integration by parts to make the gauge invariance manifest, one also finds
\begin{align}
    \epsilon^{\mu \nu \rho \sigma \lambda}\bigl(R_{\mu \nu a b}\left(\omega_{+}\right)-G_{\mu \nu} G_{a b}\bigr)\bigl(R_{\rho \sigma}^{\ \ \ a b}\left(\omega_{+}\right)-G_{\rho \sigma} G^{a b}\bigr) C_{\lambda}\kern-12em&\nn\\
    &\to\epsilon^{\mu\nu\rho\sigma\lambda}R_{\mu\nu ab}R_{\rho\sigma}{}^{ab}C_\lambda+\epsilon^{\mu\rho\sigma\lambda\gamma}G_{\mu\lambda}G_\sigma{}^\delta\nabla_\rho G_{\gamma\delta}\nn\\
    &\quad +2RG^2-8R^{\mu\nu}G_{\mu\sigma}G_\nu{}^\sigma+2R^{\mu\nu\rho\sigma}G_{\mu\nu}G_{\rho\sigma}+G^4\nn\\
    &\quad -2\epsilon^{\mu\nu\rho\sigma\lambda}R_{\mu\nu ab}G_{\rho\sigma}G^{ab}C_\lambda-4\epsilon^{\mu\nu\rho\sigma\lambda}\nabla_\mu H_{\nu ab}G_{\rho\sigma}G^{ab}C_\lambda\nn\\
    &\quad -4\epsilon^{\mu\nu\rho\sigma\lambda}H_{\mu ac}H_{\nu}{}^c{}_{b}G_{\rho\sigma}G^{ab}C_\lambda.
\end{align}
The last three terms in this expression look slightly concerning, but they will be precisely cancelled by those in
\begin{align}
    \epsilon^{\mu \nu \rho \sigma \lambda} B_{\rho \sigma}\bigl(R_{\mu \nu a b}\left(\omega_{+}\right)-G_{\mu \nu} G_{a b}\bigr) \nabla_{\lambda}\left(\omega_{+}\right) G^{a b}\kern-10em&\nn\\
    &\to -R_{\mu\nu\rho\sigma}G^{\mu\nu}G^{\rho\sigma}-\frac{1}{4}\epsilon_{\mu\nu\rho\sigma\lambda}G^{\mu\nu}G^{\rho\sigma}\nabla_{\tau}G^{\tau\lambda}+\frac{1}{2}G^4+\frac{1}{4}\qty(G^2)^2\nn\\
    &\quad +\frac{1}{2}\epsilon^{\mu\nu\rho\sigma\lambda}R_{\mu\nu ab}G_{\rho\sigma}G^{ab}C_\lambda+\epsilon^{\mu\nu\rho\sigma\lambda}\nabla_\mu H_{\nu ab}G_{\rho\sigma}G^{ab}C_\lambda\nn\\
    &\quad +\epsilon^{\mu\nu\rho\sigma\lambda}H_{\mu ac}H_{\nu}{}^c{}_{b}G_{\rho\sigma}G^{ab}C_\lambda.
\end{align}
Finally, we just need
\begin{align}
    \qty(\nabla_{\mu}\left(\omega_{+}\right) G_{a b})^2=&\qty(\nabla G)^2-\epsilon_{\mu\nu\rho\lambda\delta}G^{\lambda\delta}G^\rho{}_\beta\nabla^\mu G^{\nu\beta}-\frac{1}{4}\qty(G^2)^2+\frac{1}{2}G^4.
\end{align}

Using the above expressions along with \eqref{eq:identities1}, and making use of appropriate field redefinitions, we get
\begin{align}
    e^{-1}\cL_{(\text{Riem})^2}&=-\frac{1}{2}\mathcal I_{\text{GB}}+\frac{1}{2}R_{\mu\nu\rho\sigma}F^{\mu\nu}F^{\rho\sigma}+\frac{1}{8}(F^2)^2-\frac{11}{12}F^4-\frac{1}{2\sqrt{6}}\epsilon^{\mu\nu\rho\sigma\lambda}R_{\mu\nu ab}R_{\rho\sigma}{}^{ab}A_\lambda\nn\\
    &\quad-\frac{5}{3}g^2 F^2-60g^4.
\label{eq:invariant2}
\end{align}
Thus, we immediately see that
\begin{equation}
    \cL_{(\text{Riem})^2}+2\cL_{C^2+\frac{1}{6}R^2}=-\frac{1}{3}g^2 F^2-20 g^4,
\label{eq:actionRelation}
\end{equation}
which vanishes in the ungauged limit.  We note here also that the supersymmetrized Gauss-Bonnet Lagrangian may be written
\begin{equation}
    \cL_\text{GB}^{\text{susy}}=\cL_{(\text{Riem})^2}+3\cL_{C^2+\tfrac{1}{6}R^2},
\label{eq:GB}
\end{equation}
which we will make use of as an analogue to the 4D case \cite{Bobev:2021oku,Bobev:2020egg}.

\subsection{The action corresponding to $R^2$}
Finally, the combination of the $R^2$ Lagrangian with the usual two-derivative action has been found in \cite{Ozkan:2013nwa} (in the language of very special geometry for supergravity coupled to vector multiplets) to be
\begin{align}
    e^{-1}\cL_{R+\alpha R^2}&=\frac{1}{4}(\mathcal{C}+3) R+\frac{2}{3}(104 \mathcal{C}-8) T^{2}+8(\mathcal{C}-1) D+\frac{3}{2} C_{I J K} \rho^{I} F_{a b}^{J} F^{a b K}\nn\\
    &\quad+3 C_{I J K} \rho^{I} \partial_{\mu} \rho^{J} \partial^{\mu} \rho^{K}-24 C_{I J K} \rho^{I} \rho^{J} F_{a b}^{K} T^{a b}+\frac{1}{4} \epsilon^{a b c d e} C_{I J K} A_{a}^{I} F_{b c}^{J} F_{d e}^{K}\nn\\
    &\quad+a_{I} \rho^{I}\left(\frac{9}{64} R^{2}-3 D R-2 R T^{2}+16 D^{2}+\frac{64}{3} D T^{2}+\frac{64}{9}\left(T^{2}\right)^{2}\right),
\end{align}
where $\cC$ is an auxiliary field and $a_I$ parametrizes the $R^2$ corrections. The $D$ equation of motion gives
\begin{equation}
    \cC=1-\frac{1}{8}a_I\rho^I\qty(-3R+32D+\frac{64}{3}T^2).
\end{equation}
Substituting this back in and truncating out the vector multiplets by taking $\rho^I$ to be constant gives
\begin{align}
    e^{-1}\cL_{R+\alpha R^2}&=R+64T^2\frac{3}{2} C_{I J K} \rho^{I} F_{a b}^{J} F^{a b K}\nn\\
    &\quad-24 C_{I J K} \rho^{I} \rho^{J} F_{a b}^{K} T^{a b}+\frac{1}{4} \epsilon^{a b c d e} C_{I J K} A_{a}^{I} F_{b c}^{J} F_{d e}^{K}\nn\\
    &\quad+a_I\rho^I\qty(\frac{15}{64}R^2-DR+\frac{70}{3}RT^2-16D^2-\frac{832}{3}DT^2-\frac{1600}{9}(T^2)^2).
\end{align}
Using the fact that
\begin{subequations}
\begin{align}
    D&=-\frac{1}{32}R+\frac{2}{9}T^2,\\
    R&=\frac{64}{9}T^2-20g^2,
\end{align}
\end{subequations}
we finally get that, after field redefinitions,
\begin{equation}
    e^{-1}\cL_{R^2}=-\frac{205}{24}g^2F^2+100 g^4.
\label{eq:R2os}
\end{equation}

\subsection{The complete four-derivative action}

Given the three invariants, a generic four-derivative action in minimal 5D supergravity can be parametrized by three coefficients
\begin{equation}
    S_{\text{HD}}=S_{2\partial}+\alpha_1S_{C^2+\tfrac{1}{6}R^2}+\alpha_2 S_{(\text{Riem})^2}+\alpha_3 S_{R^2},
\label{eq:gen4ds}
\end{equation}
where the $\alpha_i$ are taken to be small such that the higher-derivative expansion is well-defined. By making use \eqref{eq:actionRelation}, \eqref{eq:GB} and \eqref{eq:R2os}, this is equivalent (at the four-derivative level) up to field redefinitions to
\begin{equation}
    S_{\text{HD}}=S_{2\partial}+(\alpha_1-2\alpha_2)S_\text{GB}^\text{susy}+g^2\int \qty[\frac{8\alpha_1-24\alpha_2-205\alpha_3}{12}F\land\star F+(\alpha_1-2\alpha_2+5\alpha_3)20g^2\star 1].
\end{equation}
We would like to make the last portion of this expression manifestly proportional to $S_{2\partial}$. To accomplish this, we perform one additional redefinition
\begin{subequations}
\begin{align}
    A&\to \qty(1+g^2 b_1)A,\\
    g^2&\to\qty(1+g^2b_2)g^2,
\end{align}
\label{eq:Ag2}%
\end{subequations}
where we assume $b_i\sim\cO(\alpha_j)$, so that we may ignore higher-order terms that appear. This field redefinition will then only pick up terms from the two-derivative action
\begin{align}
    S_\text{HD}=&S_{2\partial}+(\alpha_1-2\alpha_2)S_\text{GB}^\text{susy}+g^2\int \left[\qty(\frac{8\alpha_1-24\alpha_2-205\alpha_3}{12}-b_1)F\land\star F\right.\nn\\
    &\left.+\qty(20\alpha_1-40\alpha_2+100\alpha_3+12b_2)g^2\star1-\frac{3b_1}{3\sqrt{3}}F\land F\land A\right].
\end{align}
We may then make use of the two-derivative equations of motion to rewrite this as
\begin{align}
    S_\text{HD}=&S_{2\partial}+(\alpha_1-2\alpha_2)S_\text{GB}^\text{susy}+3b_1 g^2\int\left\{\qty[\frac{1}{3b_1}\qty(\frac{8\alpha_1-24\alpha_2-205\alpha_3}{12}-b_1)-\frac{1}{3}]F\land\star F\right.\nn\\
    &\left.+R\star1+\qty(\frac{20\alpha_1-40\alpha_2+100\alpha_3+12b_2}{3b_1}+20)g^2\star1-\frac{1}{3\sqrt{3}}F\land F\land A\right\}.
\end{align}
We must then choose $b_1$ and $b_2$ such that
\begin{subequations}
\begin{align}
    \frac{1}{3b_1}\qty(\frac{8\alpha_1-24\alpha_2-205\alpha_3}{12}-b_1)-\frac{1}{3}&=-\frac{1}{2},\\
    \frac{20\alpha_1-40\alpha_2+100\alpha_3+12b_2}{3b_1}+20&=12.
\end{align}
\end{subequations}
This is solved by
\begin{subequations}
\begin{align}
    b_1&=\frac{1}{6}\qty(8\alpha_1-24\alpha_2-205\alpha_3),\\
    b_2&=\frac{1}{6}\qty(-21\alpha_1+68\alpha_2+360\alpha_3),
\end{align}
\end{subequations}
which finally yields
\begin{equation}
    S_\text{HD}=\qty(1+g^2\frac{8\alpha_1-24\alpha_2-205\alpha_3}{2})S_{2\partial}+(\alpha_1-2\alpha_2)S_\text{GB}^\text{susy}.
\label{eq:HDaction}
\end{equation}
This is in direct analogy to the 4D case \cite{Bobev:2020egg,Bobev:2021oku}.  By use of field redefinitions, we have been able to rewrite the general four-derivative corrected action, (\ref{eq:gen4ds}), in the canonical basis of (\ref{eq:param4d}).  However, it is important to recall that this is a perturbative result valid only to linear order in the $\alpha_i$ coefficients.  From an effective field theory point of view, this is sufficient for most purposes, including computing the on-shell value of the action, as physical observables are invariant under field redefinitions.  However, straightforward use of (\ref{eq:HDaction}) is not valid for off-shell quantities unless the effects of the field redefinitions, (\ref{eq:dgdA}) and (\ref{eq:Ag2}), are fully accounted for.

This result has a strong implication. Just as in the 4D case, at the four-derivative level, the effective supergravity action is completely parametrized by two quantities, $\{S_{2\partial},S_{\mathrm{GB}}\}$, related to the geometry and three independent quantities, $\{\alpha_i,g\}$, related to the particular theory%
\footnote{While there are three $\alpha_i$ parameters, they only enter in two independent combinations in (\ref{eq:HDaction}).}.
Unlike the 4D case, however, $S_{\mathrm{GB}}^\text{susy}$ is not topological, so there can be a potentially richer structure of background geometry dependence in the AdS$_5$/CFT$_4$ setup.

We now give the explicit form of the supersymmetrized Gauss-Bonnet action $S_{\mathrm{GB}}^\text{susy}$ introduced in (\ref{eq:GB}).  Following \cite{Myers:2009ij}, we parametrize the 5D four-derivative Lagrangian as%
\footnote{This parametrization differs from \cite{Myers:2009ij} in $c_1$ and $c_2$, but is chosen for easier comparison with \cite{Cremonini:2020smy}.}
\begin{equation}
    e^{-1}\cL_{4\partial}=c_1\mathcal I_\text{GB}+c_2 C_{\mu\nu\rho\sigma}F^{\mu\nu}F^{\rho\sigma}+c_3 (F^2)^2+c_4 F^4+c_5\epsilon^{\mu\nu\rho\sigma\lambda}R_{\mu\nu ab}R_{\rho\sigma}^{\ \ \ ab}A_{\lambda}.\label{eq:4derivPart}
\end{equation}
Supersymmetry fixes this correction in terms of a single overall coefficient.  Using the definition of the five-dimensional Weyl tensor, we can make the substitution
\begin{equation}
    R_{\mu\nu\rho\sigma}F^{\mu\nu}F^{\rho\sigma}=C_{\mu\nu\rho\sigma}F^{\mu\nu}F^{\rho\sigma}+\frac{4}{3}F^4-\frac{1}{4}\qty(F^2)^2-2g^2F^2,
\end{equation}
obtained in Appendix \ref{app:simplifications} in the expressions (\ref{eq:invariant1}) and (\ref{eq:invariant2}).  Taking the Gauss-Bonnet combination, (\ref{eq:GB}), then gives
\begin{equation}
    S_{\mathrm{GB}}=\int d^5x\sqrt{-g}\qty[\mathcal I_\text{GB}-\frac{1}{2} C_{\mu\nu\rho\sigma}F^{\mu\nu}F^{\rho\sigma}+\frac{1}{8} F^4+\frac{1}{2\sqrt{3}}\epsilon^{\mu\nu\rho\sigma\lambda}R_{\mu\nu ab}R_{\rho\sigma}{}^{ab}A_{\lambda}],\label{eq:GBaction}
\end{equation}
which corresponds to
\begin{equation}
    c_1=-2 c_2=8 c_4=2\sqrt{3} c_5,\qquad c_3=0.
\label{eq:coeffs}
\end{equation}
As mentioned in the introduction, this is in conflict with some prior results \cite{Myers:2009ij,Cremonini:2020smy, Cremonini:2021jlmt2}.  However, support for the present result can be obtained from investigating supersymmetric BMPV black holes, which we turn to next.

\section{An application: the BMPV solution}\label{sec:BMPV}

As we have just seen, our main result, \eqref{eq:coeffs}, disagrees with several results in the literature. Thus, we would like to establish some evidence for the present coefficients. We note that the distinction is subtle, as the previously obtained four-derivative action of \cite{Cremonini:2020smy} differs only by a term proportional to $(F^2)^2-2F^4$, which will vanish for purely electric or purely magnetic solutions. This is because a purely electric black hole will have $F_{tr}$ as the only non-vanishing component of the field strength. One then has
\begin{subequations}
\begin{align}
    F^2&=g^{\mu\rho}g^{\nu\sigma}F_{\mu\nu}F_{\rho\sigma}=2g^{tt}g^{rr}(F_{tr})^2,\\
    F^4&=2(g^{tt})^2(g^{rr})^2F_{tr}F_{rt}F_{tr}F_{rt}=2\qty(g^{tt}g^{rr}(F_{tr})^2)^2.
\end{align}
\end{subequations}
So then we see that the combination $(F^2)^2-2F^4$ vanishes.  Another way to see is is to note that this combination can be written as
\begin{equation}
    \epsilon\epsilon F^4\equiv\epsilon_{\alpha\mu_1\mu_2\mu_3\mu_4}\epsilon^{\alpha\nu_1\nu_2\nu_3\nu_4}F^{\mu_1}{}_{\nu_1}F^{\mu_2}{}_{\nu_2}F^{\mu_3}{}_{\nu_3}F^{\mu_4}{}_{\nu_4}=-3((F^2)^2-2F^4),
\label{eq:eeF4}
\end{equation}
where the overall minus sign arises since we are using the signature $(-,+,+,+,+)$.  This also vanishes for purely magnetic objects, as the combination $(F^2)^2-2F^4$ is only sensitive to solutions where both electric and magnetic fields are present. So, to see the distinction, we must either consider a dyonic solution or a charged, rotating solution. Hence we turn to the BMPV solution \cite{Breckenridge:1996mpv}, which is a rotating black hole in five dimensions.

The BMPV solution is an asymptotically Minkowski solution, which corresponds to ungauged supergravity (or, equivalently, the $g\to0$ limit); we consider an asymptotically flat solution here as we can avoid worrying about subtleties having to do with extra factors of the two-derivative action, which simplifies the analysis. The BMPV solution is as follows
\begin{subequations}
\begin{align}
    \dd{s^2}&=-f(r)^{-2}\qty[\dd{t}+\frac{\mu\omega}{r^2}(\sin^2\theta\dd{\phi}-\cos^2\theta\dd{\psi})]^2\nn\\
    &\quad+f(r)\qty[\dd{r}^2+
    r^2\left(\sin^2\theta \dd{\phi}^2 +\cos^2\theta\dd{\psi}^2+\dd{\theta}^2\right)],\\
    A&=\sqrt{3}f(r)^{-1}\qty[\dd{t}+\frac{\mu\omega}{r^2}(\sin^2\theta\dd{\phi}-\cos^2\theta\dd{\psi})],
\end{align}
\end{subequations}
where
\begin{equation}
    f(r)=1+\frac{\mu}{r^2}.
\end{equation}
This solution depends on two parameters, $\mu$ and $\omega$, and describes a charged, spinning black hole with ADM mass
\begin{equation}
    M=\frac{3\pi}{4}\mu,
\end{equation}
two equal magnitude angular momenta in the independent planes defined by $\phi$, $\psi$,
\begin{subequations}
\begin{equation}
    J_{\phi}=\frac{\pi}{4}\mu\omega,
\end{equation}
\begin{equation}
    J_\psi=-\frac{\pi}{4}\mu\omega,
\end{equation}
\end{subequations}
and electric charge
\begin{equation}
    Q=\frac{\sqrt{3}\pi}{2}\mu.
\end{equation}
Being a supersymmetric solution, the BMPV solution satisfies the BPS equation%
\footnote{In AdS, the BPS condition includes the angular momenta, $M=\fft{\sqrt3}2|Q|+(|J_1|+|J_2|)/L$, which is another reason why the asymptotically flat case is simpler to study.}
\begin{equation}
    M=\frac{\sqrt{3}}{2}|Q|.\label{eq:BPS}
\end{equation}
The key point is that we expect \eqref{eq:BPS} to hold even after four-derivative corrections are taken into account, as the system ought to remain supersymmetric.

There is a simple argument that the four-derivative terms \eqref{eq:GBaction} do not modify the charge of the BMPV solution%
\footnote{The charge may be modified by the factor that appears in front of the two-derivative action after going on-shell, analagous to the four-dimensional case \cite{Bobev:2020egg,Bobev:2021oku}, but we focus only on the four-derivative part here.}.
Heuristically, the four-derivative terms look like two-derivative terms squared, so the equations of motion should pick up terms that are more suppressed as $r$ becomes large\footnote{Note this argument only works for asymptotically Minkowski space, where we expect solutions to fall off at infinity. In AdS, we know that objects such as the Riemann tensor will go to a constant (with respect to $r$) rather than disappearing.}. Thus, one expects the corrections to $A$ to be subleading in $r$. The charge is computed by integrating over an $S^3$ at $r\to\infty$, so we expect to only pick up the ${1}/{r^3}$ terms in $F$. For example, one might worry about terms of the form $F^3$, Weyl$\cdot F$, or Riem$^2$ contributing, but these must fall off faster than $1/r^3$ since $F$ falls off like $1/r^3$ and Riemann must fall off like $1/r^2$ in an asymptotically flat background%
\footnote{Any black hole solution with a spatially localized horizon should look more-or-less pointlike very far away (near spatial infinity), and hence the Riemann tensor should fall off no slower than for Schwarzschild, which falls off as $1/r^2$.}.
Hence, the subleading corrections from the four-derivative terms should not contribute, and the charge should not change when one introduces higher-derivative corrections. Requiring that \eqref{eq:BPS} hold in the four-derivative case then immediately implies that the mass must not shift when one introduces four-derivative corrections.

The most direct way to access the mass would simply be to find the four-derivative corrected solution and compute the ADM mass. However, this is difficult to do, so we will use a slightly more indirect approach. The on-shell action is naturally identified with the (classical) free energy
\begin{equation}
    I_{\text{HD}}=F.
\end{equation}
We have the standard thermodynamic relation
\begin{equation}
    F=U-T\cS,
\end{equation}
where $U$ is the internal energy, $T$ is the temperature, and $\cS$ is the entropy. For a black hole, the internal energy should just be the mass $M$. Moreover, since the BMPV solution is extremal, it has zero temperature, so we are left with
\begin{equation}
    I_\text{HD}=M.
\end{equation}
Noting that the mass should not change (since we need to maintain the BPS condition, and the charge is unchanged), we end up with
\begin{equation}
    \Delta I:=I_\text{HD}-I_{2\partial}=\Delta M=0.
\end{equation}

In principle, we should evaluate the action on the four-derivative solution; however, it will give the same result as evaluating it on the two-derivative solution. The standard argument is as follows. We write the four-derivative solution as a two-derivative piece plus some perturbing correction (using $\Phi$ schematically for all the fields)
\begin{equation}
    \Phi=\Phi_0+\alpha\delta\Phi+\cO(\alpha^2),
\end{equation}
where $\Phi_0$ is simply the BMPV solution in this case. Then the action is
\begin{align}
    S_\text{HD}[\Phi]&=S_{2\partial}[\Phi_0+\alpha\delta\Phi]+\alpha S_{4\partial}[\Phi_0+\alpha\delta\Phi]\nn\\
    &=S_{2\partial}[\Phi_0]+\alpha\frac{\delta S_{2\partial}}{\delta \Phi}\Big\vert_{\Phi_0}\delta\Phi+\alpha S_{4\partial}[\Phi_0]+\cO(\alpha^2)\nn\\
    &=S_\text{HD}[\Phi_0]+\cO(\alpha^2),
\end{align}
where we have used the fact that ${\delta S_{2\partial}}/{\delta \Phi}\big\vert_{\Phi_0}$ vanishes by the equations of motion.

Thus, we need only evaluate the four-derivative part of the action \eqref{eq:4derivPart} on the two-derivative solution. We will do this for generic coefficients $c_i$, and show that this necessarily leads to \eqref{eq:coeffs}. Note that the BMPV solution has nice asymptotics at infinity, so we do not need to introduce any four-derivative Gibbons-Hawking terms or boundary counterterms to remove divergences.

The Gauss-Bonnet action can be evaluated simply to be
\begin{equation}
    \int\dd[5]{x}e\mathcal I_\text{GB}=\frac{4\pi^2}{5}\frac{5\omega^4+2\omega^2\mu-15\mu^2}{\mu^2}\mathrm{vol}(\bR),
\end{equation}
where $\mathrm{vol}(\bR)$ is the (formally infinite) factor from doing the $t$ integration. The Weyl tensor contracted with graviphoton field strengths gives
\begin{equation}
    \int\dd[5]{x}eC_{\mu\nu\rho\sigma}F^{\mu\nu}F^{\rho\sigma}=-\frac{2\pi^2}{5}\frac{-160\omega^4+96\omega^2\mu+15\mu^2}{\mu^2}\mathrm{vol}(\bR).
\end{equation}
The two quartic field strength terms yield
\begin{align}
    \int\dd[5]{x}e(F^2)^2&=\frac{48\pi^2}{5}\frac{40\omega^4-48\omega^2\mu+15\mu^2}{\mu^2}\mathrm{vol}(\bR),\nn\\
    \int\dd[5]{x}eF^4&=\frac{24\pi^2}{5}\frac{20\omega^4-24\omega^2\mu+15\mu^2}{\mu^2}\mathrm{vol}(\bR).
\end{align}
Finally, the mixed Chern-Simons term gives
\begin{equation}
    \int\dd[5]{x}e\epsilon^{\mu\nu\rho\sigma\lambda}R_{\mu\nu ab}R_{\rho\sigma}{}^{ab}A_\lambda=\frac{32\sqrt{3}\pi^2}{5}\frac{5\omega^4-2\omega^2\mu}{\mu^2}\mathrm{vol}(\bR).
\end{equation}
Putting these terms together gives the requirement that
\begin{align}
    0=&\frac{2\pi^2\mathrm{vol}(\bR)}{5\mu^2}\left[10(c_1+16c_2+96c_3+24c_4+8\sqrt{3}c_5)\omega^4\right.\nn\\
    &\left.-4(-c_1+24c_2+288c_3+72c_4+8\sqrt{3}c_5)\omega^2\mu-15(2c_1+c_2-24c_3-12c_4)\mu^2\right].
\end{align}
Demanding that the $c_i$ coefficients be independent of the solution parameters $\mu$, $\omega$ then requires that we individually set
\begin{align}
    c_1+16c_2+96c_3+24c_4+8\sqrt{3}c_5&=0,\nn\\
    -c_1+24c_2+288c_3+72c_4+8\sqrt{3}c_5&=0,\nn\\
    2c_1+c_2-24c_3-12c_4&=0.
\end{align}
As we have five coefficients $c_i$ and only two parameters to vary, the solution is not unique.  Solving for the latter $c_i$ in terms of $c_1$ and $c_2$ gives
\begin{align}
    c_3&=-\frac{1}{16}(c_1+2c_2),\nn\\
    c_4&=\frac{1}{24}(7c_1+8c_2),\nn\\
    c_5&=-\frac{\sqrt{3}}{12}(c_1+6c_2).
\end{align}
Since $c_1$ just controls the overall coefficient of the four-derivative action, this is really a one parameter family of solutions. However, there is no ambiguity in the literature as to $c_2$ or $c_5$, so demanding that $c_2=-\frac{1}{2}c_1$ or that $c_5=\frac{1}{2\sqrt{3}}c_1$ immediately gives us
\begin{equation}
    c_2=-\frac{1}{2}c_1,\ c_3=0,\ c_4=\frac{1}{8}c_1, \ c_5=\frac{1}{2\sqrt{3}}c_1,
\end{equation}
in perfect agreement with the present result, \eqref{eq:coeffs}.

\section{Discussion}\label{sec:discussion}

We have shown that the three possible supersymmetric four-derivative terms that we can add to the 5D $\mathcal N=2$ supergravity action reduce after field redefinitions to a single four-derivative superinvariant as well as factors of the original two-derivative action. This, in turn, implied that there is a unique four-derivative piece of the action, up to an overall factor. We checked this explicitly in the case of the BMPV black hole and found excellent agreement.

Of particular note, we found $c_i$ coefficients that disagree with a number of results in the literature \cite{Cremonini:2008tw,Myers:2009ij,Cremonini:2020smy,Cremonini:2021jlmt2}. However, while we believe the particular $c_i$ determined previously are incorrect, the results which they are used to derive are still generally valid as they are predominantly applied to non-rotating, non-dyonic solutions, for which the discrepancy, in the form of $(F^2)^2-2F^4$, vanishes. As noted in (\ref{eq:eeF4}), this conflict with the previous result is in the $c_3$ and $c_4$ coefficients, and corresponds to a difference in the four-derivative Lagrangians
\begin{equation}
    e^{-1}\mathcal{L}_{4\partial}^{\mathrm{here}}=e^{-1}\mathcal{L}_{4\partial}^{\mathrm{previous}}-\fft1{18}\epsilon_{\alpha\mu_1\mu_2\mu_3\mu_4}\epsilon^{\alpha\nu_1\nu_2\nu_3\nu_4}F^{\mu_1}{}_{\nu_1}F^{\mu_2}{}_{\nu_2}F^{\mu_3}{}_{\nu_3}F^{\mu_4}{}_{\nu_4}.
\end{equation}
This suggests that the previous determination of the $c_i$ had an issue when translating the conventions of \cite{Hanaki:2006pj} to that of \cite{Cremonini:2008tw,Myers:2009ij}.  In particular, $\epsilon\epsilon=+5!$ or $\epsilon\epsilon=-5!$ is signature dependent, and an incorrect sign choice may have arisen when switching conventions.

With that being said, there are still outstanding issues. A natural solution to look at is the Gutowski-Reall solution \cite{Gutowski:2004r}, which is a one-parameter family of charged, spinning, supersymmetric black holes in AdS$_5$. Na\"ively, one expects that the four-derivative correction to the on-shell action should vanish%
\footnote{The stringy eight-derivative corrections to the Gutowski-Reall solution were recently explored in \cite{Melo:2007} where the shift to the action was indeed shown to vanish in the BPS case.};
however, after appropriate holographic renormalization, we seem to find that the four-derivative correction to the action does not vanish, beyond what we expect from the renormalization of the AdS radius (see Appendix \ref{app:GutowskiReall} for some technical details). There is a set of $c_i$ coefficients such that the four-derivative action vanishes, but this requires we have either $c_2\neq-\frac{1}{2}c_1$ or $c_5\neq\frac{1}{2\sqrt{3}}c_1$, which seems to be in conflict with current results in the literature. The alternative, however, is that there is a non-zero shift in the mass of the black hole. In order to preserve the BPS relation, this would require a renormalization of either the charge (which seems unlikely given the arguments presented in Section \ref{sec:BMPV}) or of the angular momentum. However, a shift to the mass seems unlikely since it has been shown that there are no corrections when the solution is uplifted to tree-level $\alpha'^3$-corrected type IIB supergravity \cite{Melo:2007}.

Additional puzzles come from some asymptotically flat solutions like supersymmetric black rings \cite{Elvang:2004emr} and four-dimensional dyonic STU black holes lifted to five dimensions \cite{Cremonini:2021jlmt2}. Of particular note is that the black ring reduces to the BMPV solution in the limit that the ring radius goes to zero. However, both of these solutions again have non-vanishing four-derivative action. The black ring solution has no set of coefficients $c_i$ such that the four-derivative action vanishes for all choices of solution parameters; however, it does in the BMPV limit. On the other hand, the lift of the dyonic 4D STU black hole has a unique set of $c_i$ coefficients making the four-derivative action vanish with $c_2=-\frac{1}{2}c_1$ and $c_5=\frac{1}{2\sqrt{3}}c_1$, but otherwise unrelated to other coefficients in the literature.

In the future, we would like to resolve these open puzzles regarding non-vanishing four-derivative actions of BPS solutions. Moreover, we believe it would be fruitful to dimensionally reduce to four dimensions, which would give us $\cN=2$ supergravity coupled to a vector multiplet, which is a truncation of the 4D STU model; we could then compare this with the results of \cite{Bobev:2021oku}. We would also like to extend the results of this note to the 5D STU model, or more generally $\cN=2$, $D=5$ supergravity coupled to vector multiplets. Finally, although motivated by holography and subleading corrections in supersymmetric partition functions, we have yet to actually explore this avenue, which we believe will be a worthwhile extension of our results.

\section*{Acknowledgements}
We wish to thank S.\ Cremonini, M. David, C.R.T. Jones, B.\ McPeak, Y.\ Tang and C. Uhlemann for useful discussions. This work was supported in part by the U.S. Department of Energy under grant DE-SC0007859.

\appendix
\section{Equations of Motion Simplifications}\label{app:simplifications}

Here we show a number of helpful simplifications that we use throughout Section \ref{sec:sugra}. All these simplifications make use of the two-derivative equations of motion, and can be considered as perturbative field redefinitions.  Some also use integration by parts, and hence are only valid inside the action.

We start by evaluating $(\nabla_\nu F^{\nu\mu})^2$ using the two-derivative equations of motion.
\begin{align}
    (\nabla_\nu F^{\nu\mu})^2=&\frac{1}{24}\epsilon^{\mu\nu\rho\sigma\lambda}F_{\nu\rho}F_{\sigma\lambda}\epsilon_{\mu\alpha\beta\gamma\delta}F^{\alpha\beta}F^{\gamma\delta}\nn\\
    =&-\frac{1}{24}\delta^{\nu\rho\sigma\lambda}_{\alpha\beta\gamma\delta}F_{\nu\rho}F_{\sigma\lambda}F^{\alpha\beta}F^{\gamma\delta}\nn\\
    =&-F_{\nu\rho}F_{\sigma\lambda}F^{[\nu\rho}F^{\sigma\lambda]}\nn\\
    =&-\frac{1}{3}(F^2)^2+\frac{2}{3}F^4.
\label{eq:simp1}
\end{align}
We quickly remark that the two-derivative equations of motion \eqref{eq:eoms} imply that
\begin{subequations}
\begin{align}
    R_{\mu\nu}F^{\mu\sigma}F^\nu{}_\sigma&=F^4-\frac{1}{6}\qty(F^2)^2-4g^2F^2,\\
    R&=\frac{1}{6}F^2-20g^2.
\end{align}
\end{subequations}
Next, we evaluate
\begin{subequations}
\begin{align}
    F^{\nu\rho}[\nabla_\mu,\nabla_\nu]F^\mu{}_\rho&=F^{\nu\rho}\qty(R^{\delta\mu}{\nu\mu}F_{\delta\rho}+R^{\delta}{}_{\rho\nu\mu}F^\mu{}_\delta)\\
    &=F^{\nu\rho}\qty(R^\delta{}_\nu F_{\delta\rho}+R_{\delta\rho\nu\mu}F^{\mu\delta})\\
    &=R^{\delta\nu}F_{\delta\rho}F_\nu{}^\rho+\frac{1}{2}R_{\nu\rho\delta\mu}F_{\mu\delta}F_{\nu\rho}
\label{eq:Bianchi2}\\
    &=F^4-\frac{1}{6}\qty(F^2)^2-\frac{1}{2}R_{\mu\nu\rho\sigma}F^{\mu\nu}F^{\rho\sigma}-4g^2F^2.
\label{eq:simp3}
\end{align}
\end{subequations}
where \eqref{eq:Bianchi2} follows from the first Bianchi identity for the Riemann tensor
\begin{equation}
    R_{[\mu\nu\rho]\sigma}=0.
\end{equation}
Now, we evaluate $(\nabla F)^2$. We recall the field strength Bianchi identity
\begin{equation}
    \nabla_{[\mu}F_{\nu\rho]}=0.
\end{equation}
This allows us to rewrite
\begin{equation}
    \nabla_\mu F_{\nu\rho}=\nabla_\nu F_{\mu\rho}+\nabla_\rho F_{\nu\mu}.\label{eq:Bianchi}
\end{equation}
Using this, we have
\begin{align}
    \nabla_\mu F_{\nu\rho}\nabla^\mu F^{\nu\rho}=&(\nabla_\nu F_{\mu\rho}+\nabla_\rho F_{\nu\mu})\nabla^\mu F^{\nu\rho}\nn\\
    =&2\nabla_\nu F_{\mu\rho}\nabla^\mu F^{\nu\rho}\nn\\
    \to&-2(\nabla_\mu\nabla_\nu F^{\mu\rho})F^{\nu}{}_\rho\nn\\
    =&-2(\nabla_\nu\nabla_\mu F^{\mu\rho})F^{\nu}{}_\rho-2([\nabla_\mu,\nabla_\nu]F^{\mu\rho})F^{\nu}{}_\rho\nn\\
    \to&\nabla_\mu F^{\mu\rho}\nabla_\nu F^{\nu\rho}-2F^{\nu\rho}[\nabla_\mu,\nabla_\nu]F^{\mu}{}_{\rho}\nn\\
    =&-\frac{1}{3}(F^2)^2-\frac{2}{3}F^4+R_{\mu\nu\rho\sigma}F^{\mu\nu}F^{\rho\sigma}+8g^2F^2,
\end{align}
where the arrows denote integration by parts, which is valid as long as we are applying this formula inside an integral. The last line follows from the equations of motion and \eqref{eq:simp1} and \eqref{eq:simp3}.

Now, we wish to evaluate
\begin{align}
    \epsilon_{\mu\nu\rho\sigma\lambda}F^{\mu\nu}F^{\rho\sigma}\nabla_\tau F^{\tau\lambda}&=-\frac{1}{2\sqrt{6}}\epsilon_{\mu\nu\rho\sigma\lambda}F^{\mu\nu}F^{\rho\sigma}\epsilon^{\lambda\alpha\beta\gamma\delta}F_{\alpha\beta}F_{\gamma\delta}\nn\\
    &=\frac{1}{2\sqrt{6}}\delta_{\mu\nu\rho\sigma}^{\alpha\beta\gamma\delta}F^{\mu\nu}F^{\rho\sigma}F_{\alpha\beta}F_{\gamma\delta}\nn\\
    &=\frac{12}{\sqrt{6}}F_{\mu\nu}F_{\rho\sigma}F^{[\mu\nu}F^{\rho\sigma]}\nn\\
    &=\frac{4}{\sqrt{6}}\qty[\qty(F^2)^2-2F^4].
\label{eq:simpeFFdF}
\end{align}
Finally, we compute
\begin{subequations}
\begin{align}
    \epsilon_{\mu\nu\rho\lambda\delta}F^{\lambda\delta}F^{\rho}{}_\beta\nabla^\mu F^{\nu\beta}&=\frac{1}{2}\epsilon_{\mu\nu\rho\lambda\delta}F^{\lambda\delta}F^{\rho}{}_\beta\nabla^\beta F^{\nu\mu}\label{eq:simp2}\\
    &=-\frac{1}{2}\epsilon_{\mu\nu\rho\lambda\delta}F^{\lambda\delta}F^{\rho}{}_\beta\nabla^\beta F^{\mu\nu}\\
    &=-\frac{1}{4}\epsilon_{\mu\nu\rho\lambda\delta}F^{\rho}{}_\beta\nabla^\beta \qty(F^{\mu\nu}F^{\lambda\delta})\\
    &\to\frac{1}{4}\epsilon_{\mu\nu\rho\lambda\delta}\qty(\nabla^\beta F^{\rho}{}_\beta) F^{\mu\nu}F^{\lambda\delta}\\
    &=-\frac{1}{4}\epsilon_{\mu\nu\rho\lambda\delta}F^{\mu\nu}F^{\lambda\delta}\nabla_\beta F^{\beta\rho}\\
    &=-\frac{1}{\sqrt{6}}\qty[\qty(F^2)^2-2F^4],
\label{eq:simp4}
\end{align}
\end{subequations}
where \eqref{eq:simp2} follows from the Bianchi identity \eqref{eq:Bianchi}. The last line in the computation follows from \eqref{eq:simpeFFdF}.

We can also evaluate some curvature squared terms using \eqref{eq:eoms}
\begin{equation}
    R_{\mu\nu}R^{\mu\nu}=F^4-\frac{7}{36}\qty(F^2)^2-\frac{4}{3}g^2F^2+80g^4,
\end{equation}
and also
\begin{equation}
    R^2=\frac{1}{36}\qty(F^2)^2-\frac{20}{3}g^2F^2+200g^4.
\end{equation}

One last useful formula for us is
\begin{align}
    R_{\mu\nu\rho\sigma}F^{\mu\nu}F^{\rho\sigma}&=\qty(C_{\mu\nu\rho\sigma}+\frac{4}{3}g_{\mu\rho}R_{\sigma\nu}-\frac{1}{6}Rg_{\mu\rho}g_{\sigma\nu})F^{\mu\nu}F^{\rho\sigma}\nn\\
    &=C_{\mu\nu\rho\sigma}F^{\mu\nu}F^{\rho\sigma}+\frac{4}{3}R_{\sigma\nu}F^{\rho\nu}F_\rho{}^\sigma-\frac{1}{6}RF^2\nn\\
    &=C_{\mu\nu\rho\sigma}F^{\mu\nu}F^{\rho\sigma}+\frac{4}{3}F^4-\frac{1}{4}\qty(F^2)^2-2g^2F^2.
\end{align}
%

\section{The Gutowski-Reall Black Hole}\label{app:GutowskiReall}

Here we compute the on-shell value of the parametrized four-derivative corrected action, (\ref{eq:param4d}), for the Gutowski-Reall black hole \cite{Gutowski:2004r}.  When viewed as an asymptotically AdS$_5\times S^5$ solution to IIB supergravity, it is known that the first correction occurs at the eight-derivative level.  The corrected action with curvature and the Ramond-Ramond five-form was obtained in \cite{Gross:1986iv,Paulos:2008tn}, and it was shown in \cite{Melo:2007} that the on-shell eight-derivative correction vanishes for the supersymmetric Gutowski-Reall solution.  However, here we take more of a bottom up view and introduce four-derivative corrections to the five-dimensional action as may occur in theories with reduced supersymmetry.

The Gutowski-Reall black hole \cite{Gutowski:2004r} is a solution of minimal gauged supergravity in 5D given by
\begin{subequations}
\begin{equation}
    \dd{s}^2=-U(r)\Lambda(r)^{-1}\dd{t}^2+U(r)^{-1}\dd{r}^2+\frac{r^2}{4}\qty[\qty(\sigma^{1'}_L)^2+\qty(\sigma^{2'}_L)^2+\Lambda(r)\qty(\sigma_L^{3'}-\Omega(r)\dd{t})^2],
\end{equation}
\begin{equation}
    A=\sqrt{3}\left[\left(1-\frac{R_{0}^{2}}{r^{2}}-\frac{R_{0}^{4}}{2 L^{2} r^{2}}\right) \dd{t}+\frac{\epsilon R_{0}^{4}}{4 Lr^{2}} \sigma_{L}^{3'}\right],
\end{equation}
\end{subequations}
where
\begin{subequations}
\begin{equation}
    U(r)=\qty(1-\frac{R_0^2}{r^2})\qty(1+\frac{2R_0^2}{L^2}+\frac{r^2}{L^2}),
\end{equation}
\begin{equation}
    \Lambda(r)=1+\frac{R_0^6}{L^2r^4}-\frac{R_0^8}{4L^2r^6},
\end{equation}
\begin{equation}
    \Omega(r)=\frac{2\epsilon}{L\Lambda(r)}\qty[\qty(\frac{3}{2}+\frac{R_0^2}{L^2})\frac{R_0^4}{r^4}-\qty(\frac{1}{2}+\frac{R_0^2}{4L^2})\frac{R_0^6}{r^6}],
\end{equation}
\begin{equation}
    \sigma_L^{1'}=\sin\phi\dd{\theta}-\cos\phi\sin\theta\dd{\psi},
\end{equation}
\begin{equation}
    \sigma_L^{2'}=\cos\phi\dd{\theta}+\sin\phi\sin\theta\dd{\psi},
\end{equation}
\begin{equation}
    \sigma_L^{3'}=\dd{\phi}+\cos\theta\dd{\psi},
\end{equation}
\end{subequations}
and $\epsilon^2=\pm1$. We also have that
\begin{equation}
    \theta\in[0,\pi),\ \ \phi\in\left[\frac{2\epsilon t}{L},4\pi+\frac{2\epsilon t}{L}\right),\ \ \psi\in[0,2\pi)
\end{equation}
in these coordinates. This solution corresponds to a charged spinning black hole with mass
\begin{equation}
    M=12 \pi^2 R_{0}^{2}\left(1+\frac{3 R_{0}^{2}}{2 L^{2}}+\frac{2 R_{0}^{4}}{3 L^{4}}\right),
\end{equation}
angular momenta
\begin{subequations}
\begin{align}
    J_\phi&=\frac{6 \epsilon \pi^2 R_{0}^{4}}{L}\left(1+\frac{2 R_{0}^{2}}{3 L^{2}}\right),\\
    J_\psi&=0,
\end{align}
\end{subequations}
and charge
\begin{equation}
    Q=8\sqrt{3} \pi^2 R_{0}^{2}\left(1+\frac{R_{0}^{2}}{2 L^{2}}\right).
\end{equation}
This is easily seen to satisfy the BPS equation
\begin{equation}
    M-\frac{2}{L}|J|=\frac{\sqrt{3}}{2}|Q|.
\end{equation}

In the coordinates induced from the bulk solution, the boundary metric becomes
\begin{equation}
    \dd{s}^2_\text{bdy}=-\dd{t}^2+\frac{L^2}{4}\qty(\qty(\sigma_L^{1'})^2+\qty(\sigma_L^{2'})^2+\qty(\sigma_L^{3'})^2).
\end{equation}
Hence, we see that the boundary topology is $\bR\times S^3$. As a result, we should not expect a conformal anomaly, i.e., we do not need to cancel any logarithmic divergences.

The two-derivative action \eqref{eq:2derivAction} is simple to compute. Using the standard Gibbons-Hawking term \cite{Gibbons:1977h}
\begin{equation}
    S_{2\partial}^\text{GH}=2\int\dd[4]{x}\sqrt{-h}K,
\end{equation}
and boundary counterterm
\begin{equation}
    S_{2\partial}^\text{CT}=\int\dd[4]{x}\qty(\frac{6}{L}+\frac{L}{2}\cR),
\end{equation}
to cancel the divergences, we find the holographically renormalized two-derivative action is
\begin{equation}
    I_{2\partial}^{\text{Ren}}=\frac{\pi^2T}{2L^2}\qty(-3L^4+4R_0^4).
\end{equation}

Now, we must compute five pieces of the four-derivative action
\begin{subequations}
\begin{align}
    S_1&:=\int\dd[5]{x}e\hat{R}_\text{GB}\\
    S_2&:=\int\dd[5]{x}e C_{\mu\nu\rho\sigma}F^{\mu\nu}F^{\rho\sigma}\\
    S_3&:=\int\dd[5]{x}e\qty(F^2)^2\\
    S_4&:=\int\dd[5]{x}eF^4\\
    S_5&:=\int\dd[5]{x}e\epsilon^{\mu\nu\rho\sigma\lambda}R_{\mu\nu ab}R_{\rho\sigma}^{ab}A_\lambda
\end{align}
\end{subequations}
The only term we expect to give rise to divergences is $S_1$. This can be cured with an appropriate Gibbons Hawking term \cite{Teitelboim:1987z,Myers:1987yn}
\begin{equation}
    S_1^{\text{GH}}=2\int\d[4]{x}\sqrt{-h}\qty[-\frac{2}{3}K^3+2KK_{ab}K^{ab}-\frac{4}{3}K_{ab}K^{bc}K_c{}^a-4\qty(\cR_{ab}-\frac{1}{2}\cR h_{ab})K^{ab}]
\end{equation}
where $K_{ab}$ is the extrinsic curvature, $K=h^{ab}K_{ab}$ is its trace, $\cR_{ab}$ is the induced Ricci tensor on the boundary, and $\cR$ is the induced Ricci scalar on the boundary. We also have boundary counterterms \cite{Cremonini2009ls}
\begin{equation}
    S_1^{\text{CT}}=-\int\dd[4]x\sqrt{-h}\qty(-\frac{2}{L^2}+\frac{3}{2L}\cR)
\end{equation}
per the usual holographic renormalization procedure. With this in hand, it is straightforward to compute
\begin{subequations}
\begin{align}
    I_1^\text{Ren}&=\frac{\pi^2\mathrm{vol}(\bR)}{20L^4}\qty(210L^4+408L^2R_0^2+128R_0^4),\\
    I_2&=\frac{2\pi^2\mathrm{vol}(\bR)}{5L^4}\qty(-15L^4+66L^2R_0^2+55R_0^4),\\
    I_3&=\frac{24\pi^2\mathrm{vol}(\bR)}{5L^4}\qty(30L^4+36L^2R_0^2+11R_0^4),\\
    I_4&=\frac{6\pi^2\mathrm{vol}(\bR)}{5L^4}\qty(60L^4+96L^2R_0^2+41R_0^4),\\
    I_5&=\frac{9\pi^2\mathrm{vol}(\bR)}{2L^4}\qty(72L^2R_0^2-7R_0^4).
\end{align}
\end{subequations}
Of particular note is that
\begin{equation}
    I_1^\text{Ren}-\frac{1}{2}I_2+\frac{1}{8}S_I+\frac{1}{2\sqrt{3}}S_I=\frac{9\pi^2\mathrm{vol}(\bR)}{2L^2}\qty(5L^2+8R_0^2),
\end{equation}
which is neither zero nor the two-derivative result. If one writes the four-derivative action with generic coefficients
\begin{equation}
    S_{4\partial}=c_1S_1+c_2S_2+c_3S_3+c_4S_4+c_5S_5,
\end{equation}
then we see that $I_{4\partial}=0$ requires that
\begin{subequations}
\begin{align}
    -7c_1 + 4 c_2 - 96 c_3 - 48 c_4&=0,\\
    17c_1 + 22 c_2 + 144 c_3 + 96 c_4 + 12 \sqrt{3} c_5&=0,\\
    125c_1 + 440 c_2 + 1056 c_3 + 984 c_4 - 28 \sqrt{3} c_5&=0,
\end{align}
\end{subequations}
which can be solved by
\begin{subequations}
\begin{align}
    c_2&=-\frac{151c_1+512\sqrt{3}c_5}{96},\\
    c_3&=-\frac{707c_1+2368\sqrt{3}c_5}{768},\\
    c_4&=\frac{901c_1+3296\sqrt{3}c_5}{576}.
\end{align}
\end{subequations}
If we enforce that $c_5=\frac{1}{2\sqrt{3}}c_1$, we get that
\begin{equation}
    c_2=-\frac{407}{96}c_1,\qquad c_3 =-\frac{1891}{768}c_1,\qquad c_4=\frac{2549}{576}c_1.
\end{equation}
These coefficients do not agree with any known results in the literature. Alternatively, enforcing $c_2=-\frac{1}{2}c_1$ gives
\begin{equation}
    c_3=-\frac{615}{2048},\qquad c_4=\frac{423}{1024},\qquad c_5=-\frac{103}{512\sqrt{3}},
\end{equation}
which is equally undelightful.

\bibliographystyle{JHEP}
\bibliography{cite}

\end{document}